# Joint Adaptive Modulation-Coding and Cooperative ARQ for Wireless Relay Networks

Morteza Mardani*, Jalil S. Harsini*, Farshad Lahouti*, Behrouz Eliasi**

*Wireless Multimedia Communication Lab., School of E&CE, University of Tehran
**Iran Telecommunication Research Center, PO Box 14155-3961, Tehran 14399, Iran
Emails: m.mardani@ece.ut.ac.ir, j.harsini@ece.ut.ac.ir, lahouti@ut.ac.ir, eliasi@itrc.ac.ir

*Abstract*—This paper presents a cross-layer approach to jointly design adaptive modulation and coding (AMC) at the physical layer and cooperative truncated automatic repeat request (ARQ) protocol at the data link layer. We first derive an exact closed form expression for the spectral efficiency of the proposed joint AMC-cooperative ARQ scheme. Aiming at maximizing this system performance measure, we then optimize an AMC scheme which directly satisfies a prescribed packet loss rate constraint at the data-link layer. The results indicate that utilizing cooperative ARQ as a retransmission strategy, noticeably enhances the spectral efficiency compared with the system that employs AMC alone at the physical layer. Moreover, the proposed adaptive rate cooperative ARQ scheme outperforms the fixed rate counterpart when the transmission modes at the source and relay are chosen based on the channel statistics. This in turn quantifies the possible gain achieved by joint design of AMC and ARQ in wireless relay networks.

## I. INTRODUCTION

Providing quality of service (QoS) guarantees to various communications applications is an important objective of next generation wireless networks. However, wireless links are subject to time varying fading which limits their performance. Adaptive Modulation and Coding (AMC) is a powerful technique to combat the effect of fading at the physical layer, which can enhance the spectral efficiency, dramatically with respect to non-adaptive systems [1], [2]. Still, joint design of AMC and an Automatic Repeat Request (ARQ) protocol further enhances the performance [3].

Cooperative relaying has recently emerged as a powerful spatial diversity technique for improved performance [4], [5]. An information theoretic study of the relay channel is presented in the original work of Cover and El Gamal in [6]. Since then, it has been extensively studied and the associated practical protocols have been devised. Specifically, the incremental decode and forward relaying protocol, enjoys a high spectral efficiency, owing to limiting the relay to destination retransmission only to the instances when the data received at the destination is in error [4]. The selection decode and forward protocol, still checks whether the data received at the relay is correct, prior to a possible relay to destination retransmission [4]. In this direction, a cooperative ARQ protocol is presented in [7], which benefiting from the spatial diversity of the relay channel, outperforms a traditional ARQ scheme, particularly when the source-destination channel is subject to a high temporal correlation. The authors in [8] investigate a cross-layer design for combining cooperative diversity with truncated ARQ in wireless Ad-hoc networks. Using throughput performance measure they found optimal packet length and modulation level which maximize the system throughput. Their results show that the combination of cooperative diversity with truncated ARQ considerably improves the system throughput performance, when compared to the traditional truncated ARQ.

To apply the idea of AMC to the relay channel a few works are reported in the literature, e.g. [9], [10], [11]. Assuming capacity achieving codes, the work in [9] proposes a discrete power and rate adaptation algorithm for fixed decode and forward relay protocol. In [11], a rate adaptation scheme for coded-cooperation protocols based on the channel statistics is presented, when an ARQ protocol with possibly infinite number of retransmissions is used. However, designing practical AMC schemes suitable for QoS constrained applications in wireless relay networks is still an open problem.

The main contribution of this paper is to quantify the potential spectral efficiency gain achieved by joint design of discrete-rate AMC with cooperative-ARQ, while satisfying the QoS constraints of higher layers. To this end, we take a cross-layer design approach. We first derive an exact closed form expression for the spectral efficiency of joint AMC-cooperative truncated-ARQ scheme over block fading channels. Then based on this performance measure, we propose a cross-layer design maximizing the system performance subject to a packet loss rate (PLR) constraint. Numerical results show that the proposed cross-layer design achieves considerable spectral efficiency gain with respect to AMC alone at the physical layer. Moreover, it outperforms the constant rate cooperative ARQ, when the transmission modes at the source and relay are chosen based on the channel statistics.

The rest of this paper is organized as follows. Section II describes the system model, including a proposed cooperative ARQ protocol description at the data link layer and AMC at the physical layer. In Section III, we first derive the spectral efficiency of the assumed system model, and subsequently use it to propose a cross-layer design scheme. Section IV describes the cooperative ARQ in a fixed rate scenario. Numerical results are provided in section V, while the concluding remarks are presented in sections VI.

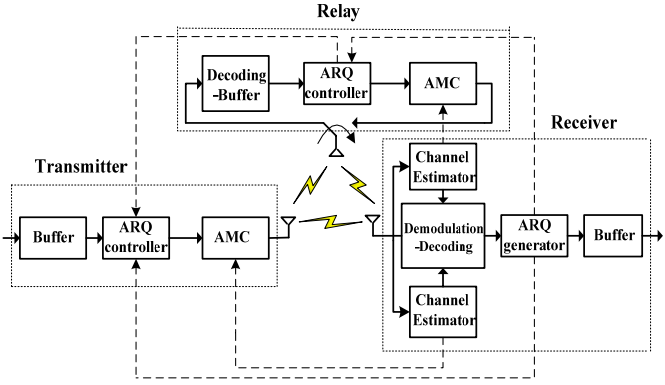

Fig. 1. System model

## II. SYSTEM MODEL
### A. System Description

As illustrated in Fig.1 we consider a wireless network composed of a source node (S), a relay node (R) and a destination node (D), where each node is equipped with a single antenna. At the source node, input packets from higher layers of protocol stack are first stored in a transmit buffer, grouped into frames, and then transmitted over the wireless channel on a frame by frame basis. We adopt the packet and frame structure as in [3], where the CRC bits of each packet facilitate perfect error detection. The considered cooperative-ARQ protocol acts as follows. First the S node transmits a data frame to both R and D nodes. Upon reception of a packet at node D, it checks the CRC for each packet and transmits either a positive or negative acknowledgement (ACK or NACK). In case, the relay receives a NACK message from the destination, and it has been able to successfully decode the corresponding packet, it retransmits the packet until it is successfully received at the destination or a maximum allowable number of retransmissions is reached. Otherwise, the S transmits a fresh packet and the above process is repeated.

### B. Channel Models and AMC Modes

For S-D, R-D and S-R channels We consider a discrete time channel model and AWGN with one-sided power spectral density $N_0$. Both S-D and R-D channels encounter Rayleigh fading with stationary and ergodic channel gains $\sqrt{h_{sd}}$ and $\sqrt{h_{rd}}$, respectively. We adopt a block fading model so that the channel gains are constant per frame and vary randomly from one frame to another [13]. Due to the suitable selection of the relay position, we assume that the S-R link is an AWGN channel with SNR $\gamma_{sr}$.

Both S and R nodes have the same constant transmit power level of $\bar{P}$ and a bandwidth $W$. The instantaneous received SNR for the S-D and R-D channels are $\gamma_1 = \bar{P}h_{sd}/N_0W$ and $\gamma_2 = \bar{P}h_{rd}/N_0W$, respectively. At the physical layer, AMC is employed for both S-D and R-D links based on their corresponding channel state information (CSI). We assume that perfect CSI is available at the destination and that the selected AMC modes are fed back to the S and R nodes reliably and without delay. The AMC can be employed for each link by dividing the entire SNR range into $N + 1$ non-overlapping consecutive intervals, denoted by $[\gamma_{i,n}, \gamma_{i,n+1})$, $i = 1,2$, $n = 0, \dots, N$, where $\gamma_{i,0} = 0$ and $\gamma_{i,N+1} = \infty$. When the instantaneous SNR $\gamma_1$ falls in the interval $[\gamma_{1,n}, \gamma_{1,n+1})$, the mode $n$ of AMC is chosen and the S transmits at the rate of $R_n$ (bits/symbol). Also, when the SNR $\gamma_2$ falls in the interval $[\gamma_{2,m}, \gamma_{2,m+1})$, the R will transmit at the rate $R_m$. No signal is transmitted when $\gamma_i \in [\gamma_{i,0}, \gamma_{i,1})$, $i = 1,2$, corresponding to the outage modes of S-D and R-D links, respectively.

As the channel gains are assumed constant over a frame, the corresponding attenuation may be compensated at the receiver, and therefore, the channel may be considered as AWGN in each frame. In order to simplify the analysis, we approximate the packet error rate (PER) for the AMC mode $n$ using the following expression [3]

$$PER_n(\gamma) \approx \begin{cases} 1, & \gamma < \Gamma_n \\ a_n \exp(-g_n\gamma), & \gamma \geq \Gamma_n \end{cases} \quad (1)$$

where the parameters $\{a_n, g_n, \Gamma_n\}$ are determined by the curve fitting to the exact PER of mode $n$.

## III. JOINT DESIGN OF AMC AND COOPERATIVE ARQ

In this section, we develop a cross-layer approach to jointly design AMC at the physical layer and cooperative ARQ at the data link layer. To guarantee a low delay and a small buffer size, we assume that the maximum number of retransmission attempts per packet at the R node is limited to $N_r$. Using a limited number of retransmissions, error free delivery of packets is not guaranteed. Therefore, if a packet is not received correctly after the relay retransmissions, it is considered as lost. In addition to the delay constraint, we also assume that the packet service to be provisioned imposes a PLR QoS constraint at the data link layer.

### A. Spectral Efficiency

Here, we derive an exact closed-form expression for the spectral efficiency of the proposed adaptive rate cooperative ARQ scheme. In [2] the spectral efficiency for an adaptive-rate scheme is defined as the average number of information bits transmitted per symbol. In this section, we develop a similar definition for the spectral efficiency of the proposed combining scheme. To this end, the following Proposition is presented.

*Proposition 1*: For the considered adaptive rate cooperative ARQ protocol, the average spectral efficiency is given by (2), where $\varepsilon_n = PER_n(\gamma_{sr})$ is the PER of AWGN S-R channel in mode $n$, $L(n, \underline{m}^l) = 1/R_n + \sum_{k=1}^{l} 1/R_{m_l}$, $\underline{m}^l = (m_1, m_2, \dots, m_l)$, $P_n^{sd} = \int_{\gamma_{1,n}}^{\gamma_{1,n+1}} p_{\gamma_1}(\gamma)d\gamma$ and

$$\overline{PER}_n^{sd} = \frac{1}{P_n^{sd}} \int_{\gamma_{1,n}}^{\gamma_{1,n+1}} PER_n(\gamma) p_{\gamma_1}(\gamma) d\gamma \quad (3)$$

The probability $P_{m_l}^{rd}$ and PER $\overline{PER}_{m_l}^{rd}$ are obtained by substituting the R-D channel parameters in $P_n^{sd}$ and $\overline{PER}_n^{sd}$.

*Proof*: consider a packet based system where each packet contains a fixed number of $N_P$ bits, and is transmitted using $L$ symbols. Each packet encounters a vector of SNR channel realizations $\underline{\gamma} = (\gamma_1, \gamma_2^1, \dots, \gamma_2^M)$ until it is received correctly

$$\eta = \sum_{n=1}^{N} \frac{(1-(1-\varepsilon_n)\overline{PER}_n^{sd})}{L(n,\underline{m}^0)} P_n^{sd} + \sum_{l=1}^{N_r-1} \sum_{n=1}^{N} \sum_{m_1=1}^{N} \cdots \sum_{m_l=1}^{N} \frac{(1-\varepsilon_n)}{L(n,\underline{m}^l)} \times \overline{PER}_n^{sd} P_n^{sd} \prod_{k=1}^{l-1} \overline{PER}_{m_k}^{rd} P_{m_k}^{rd} (1-\overline{PER}_{m_l}^{rd}) P_{m_l}^{rd}$$

$$+ \sum_{n=1}^{N} \sum_{m_1=1}^{N} \cdots \sum_{m_{N_r}=1}^{N} \frac{(1-\varepsilon_n)\overline{PER}_n^{sd} P_n^{sd} P_{m_{N_r}}^{rd}}{L(n,\underline{m}^{N_r})} \prod_{k=1}^{N_r-1} \overline{PER}_{m_k}^{rd} P_{m_k}^{rd} \qquad (2)$$

or the maximum allowable number of retransmissions is reached. Here the random variable $M \in \{1,2,..,N_r\}$ depends on the channel noise. The spectral efficiency is defined as the average number of transmitted bits per symbol, i.e.

$$\eta = E_{M,\underline{\gamma}}\left[\frac{N_P}{L}\right] \qquad (4)$$

After following some mathematical calculations the equation (4) is reduced to (2). The detailed derivation of (2) is omitted here due to lack of space and is available in [14].

*Corollary 1*: For an adaptive rate traditional ARQ scheme with a maximum number of retransmissions per packet $N_r$, when the channel gain for original transmission and retransmissions of a packet are independent, the average spectral efficiency is obtained from (2) by substituting $\varepsilon_n = 0$ and $p_{\gamma_1}(\gamma) = p_{\gamma_2}(\gamma)$.

*Proof*: The proof is straightforward from Proposition 1 by substituting $\varepsilon_n = 0$ and $p_{\gamma_1}(\gamma) = p_{\gamma_2}(\gamma)$.

### B. Optimizing the Spectral Efficiency

Based on the performance metric derived in the previous subsection, we now propose a cross-layer design for adaptive-rate cooperative-ARQ system with $N_r = 1$. Extension of the proposed analysis to the case of $N_r > 1$ is straightforward. The objective is to maximize the average system spectral efficiency subject to a prescribed error requirement as follows

$$\max_{\{\gamma_{1,n},\gamma_{2,m}\}_{n,m=1}^N} \eta \quad \text{subject to}$$
$$C: \overline{PLR} \leq P_{loss} \qquad (5)$$

where $P_{loss}$ is the target PLR, and $\overline{PLR}$ is the average system PLR. The constraint C states that the system packet loss rate is not greater than the target PLR.

*Proposition 2*: The average system PLR of the considered adaptive rate cooperative ARQ protocol is

$$\overline{PLR} = \left(\frac{\sum_{n=1}^{N}\overline{PER}_n^{sd} P_n^{sd}}{\sum_{n=1}^{N} P_n^{sd}}\right)\left(\frac{\sum_{m=1}^{N}\overline{PER}_m^{rd} P_m^{rd}}{\sum_{m=1}^{N} P_m^{rd}}\right) + \left(\frac{\sum_{n=1}^{N}\varepsilon_n\overline{PER}_n^{sd} P_n^{sd}}{\sum_{n=1}^{N} P_n^{sd}}\right)\left(1 - \frac{\sum_{m=1}^{N}\overline{PER}_m^{rd} P_m^{rd}}{\sum_{m=1}^{N} P_m^{rd}}\right) \qquad (6)$$

*Proof*: The proof is provided in appendix A.

Having described the performance measure and QoS constraints, we now consider the cross-layer design problem of interest in (5). Specifically, we propose an approach that formulates this problem into two separate designs of AMC for S-D and R-D links. In this formulation, we consider the following average PERs per mode

$$\overline{PER}_n^{sd} = P_{t,sd}, \quad n = 1,2,...,N \qquad (7)$$

and

$$\overline{PER}_m^{rd} = P_{t,rd}, \quad m = 1,2,...,N \qquad (8)$$

where $P_{t,sd}$ and $P_{t,rd}$ are target PERs. Using equations (6), (7) and (8), satisfying the PLR constraint C in (6) with equality, we have

$$P_{loss} = P_{t,sd} \times P_{t,rd} + \bar{\varepsilon} P_{t,sd}(1-P_{t,rd}) \qquad (9)$$

where,

$$\bar{\varepsilon} = \frac{\sum_{n=1}^{N}\varepsilon_n P_n^{sd}}{\sum_{n=1}^{N} P_n^{sd}} \qquad (10)$$

is the average PER over the S-R channel. The design problem is to find the optimal target PERs, $P_{t,sd}^*$ and $P_{t,rd}^*$, such that the system spectral efficiency is maximized, while satisfying the equation (9). The following algorithm describes a search method for this purpose.

Step 1) Choose $P_{t,sd} \in \mathcal{P}$, where the set $\mathcal{P}$ is

$$\mathcal{P} = \{P_{t,sd}: P_{loss} < P_{t,sd} < 1\} \qquad (10)$$

Step 2) Design AMC for the S-D link based on the given $P_{t,sd}$, and equation (7), following the approach suggested in [12].

Step 3) Compute the average PER of S-R channel using equation (10).

Step 4) Given $P_{loss}$, $\bar{\varepsilon}$, $P_{t,sd}$, using (9), we obtain

$$P_{t,rd} = \frac{P_{loss} - \bar{\varepsilon} P_{t,sd}}{P_{t,sd}(1-\bar{\varepsilon})} \qquad (11)$$

If $P_{t,rd} < 0$ go to step 7.

Step 5) Design AMC for the R-D link based on the given $P_{t,rd}$, and equation (8), following the approach suggested in [12].

Step 6) Compute $\eta(P_{t,sd})$ from (2).

Step 7) Repeating steps 1 to 6, determine the optimal $P_{t,sd}$ as follows

$$P_{t,sd}^* = \underset{P_{t,sd} \in \mathcal{P}}{\mathrm{argmax}}\ \eta(P_{t,sd}) \qquad (12)$$

Once, $P_{t,sd}^*$ and subsequently $P_{t,rd}^*$ are obtained, the design process is completed. A special case of interest is to consider a S-R channel with high SNR ($\bar{\varepsilon} \approx 0$). In this case, we have $P_{t,rd}^* \times P_{t,sd}^* = P_{loss}$. Naturally, one in general may devise more efficient design or search solutions for the last step of the algorithm.

## IV. COOPERATIVE ARQ IN FIXED RATE SCENARIO

In order to provide a benchmark for comparison, here we consider a scenario in which only the average SNR of the corresponding channels are known at the source and relay nodes and no instantaneous CSI is available. In this case the optimized transmission modes at the source and relay nodes can be selected based on the channel statistics. Consider the problem of selecting the fixed optimized rates $R_n$ and $R_m$ for the S and R nodes, respectively, aiming at maximizing the spectral efficiency subject to the average PLR constraint, i.e.,

$$\max_{n,m} \eta(n,m) \quad \text{subject to} \quad (13)$$
$$C: \overline{PLR}(n,m) \leq P_{loss}$$

In a manner similar to that proposed in proposition 1, the average spectral efficiency for this scenario is given by

$$\eta(n,m) = R_n(1-(1-\varepsilon_n)\frac{R_n}{R_n+R_m}\overline{PER}_{sd}(n)) \quad (14)$$

Using the same approach as in Appendix A, the average PLR is given by

$$\overline{PLR}(n,m) = \overline{PER}_{sd}(n)\overline{PER}_{rd}(m) \\ + \varepsilon_n \overline{PER}_{sd}(n)(1-\overline{PER}_{rd}(m)) \quad (15)$$

where

$$\overline{PER}_{sd}(n) = \int_0^\infty PER_n(\gamma) p_{\gamma_1}(\gamma) d\gamma$$
$$= \int_0^{\Gamma_n} p_{\gamma_1}(\gamma)d\gamma + \int_{\Gamma_n}^\infty a_n \exp(-g_n\gamma) p_{\gamma_1}(\gamma)d\gamma$$
$$= 1 - \exp\left(\frac{-\Gamma_n}{\bar{\gamma}_1}\right) + \frac{a_n}{1+g_n\bar{\gamma}_1} exp\left(-(g_n+\frac{1}{\bar{\gamma}_1})\Gamma_n\right)$$

We can obtain $\overline{PER}_{rd}(m)$ by substituting $n$ and $\bar{\gamma}_1$ by $m$ and $\bar{\gamma}_2$ in $\overline{PER}_{sd}(n)$. In the absence of rate adaptation, average system PLR does not satisfy the constraint C over the range of the $\bar{\gamma}_1$ and $\bar{\gamma}_2$. In fact, for each pair $(R_n,R_m)$ there is a threshold $\bar{P}_{th}$ for the transmit power of source and the relay nodes so that the constraint C is satisfied only for $\bar{P} > \bar{P}_{th}$. As in the case of adaptive rate cooperative ARQ (Section III-B), the problem in (13) can be reduced to a simpler single-variable optimization problem similar to (12).

## V. NUMERICAL RESULTS

In this section, we evaluate the performance of the proposed schemes. For both relay and source nodes we use the AMC modes of HYPERLAN/2 standard with packet length $N_P = 1080$ bits as presented in table II of [12].

We consider the scenario where the S, R and D lie along a straight line, the S-D distance is normalized to unity and the S-R distance is denoted by $d$ [9]. In this case, the channel SNRs $\gamma_1$, $\gamma_2$ are exponential random variables with mean $\bar{\gamma}_1 = \bar{P}$, , $\bar{\gamma}_2 = \bar{P}(1-d)^{-\alpha}$. Also, the SNR of AWGN S-R channel is $\bar{\gamma}_{sr} = \bar{P}d^{-\alpha}$. In our analysis we assume a path loss exponent of $\alpha = 4$ and $P_{loss} = 0.001$.

Fig. 2 depicts the average spectral efficiency versus the average SNR of S-D link ($\bar{P}$) for adaptive-rate cooperative ARQ and AMC-only schemes. We observe that the spectral efficiency of the proposed joint AMC-cooperative ARQ scheme exceeds that of AMC-only scheme by about 0.5 bits per symbol. This considerable performance gain signifies the role of retransmission by the relay.

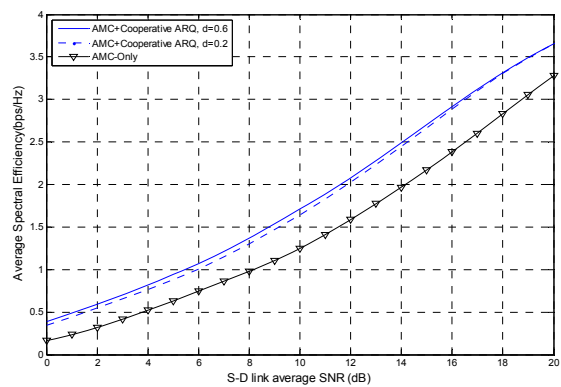

Fig. 2. Spectral efficiency vs. the S-D link SNR for both joint AMC-cooperative ARQ scheme and AMC-only scheme.

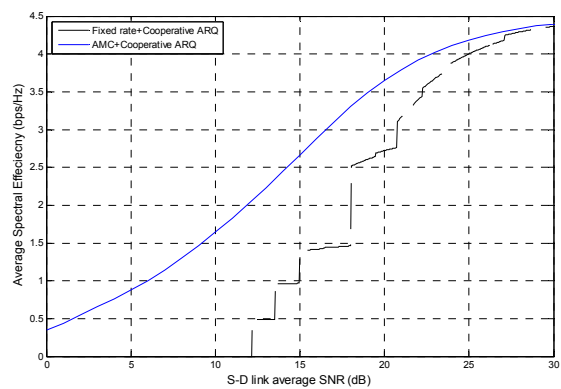

Fig. 3. Spectral efficiency vs. S-D link SNR for both adaptive rate and fixed rate cooperative ARQ schemes, $d$=0.2.

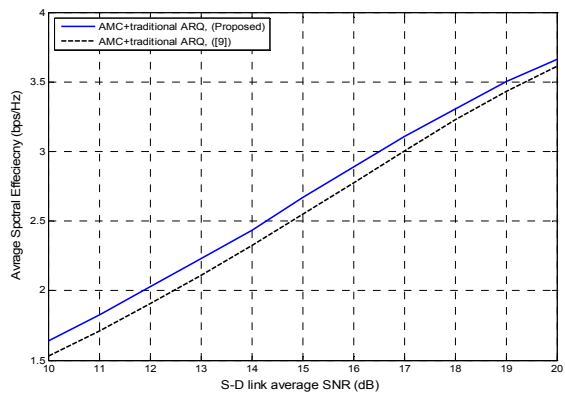

Fig. 4. Comparison between the spectral efficiency of proposed joint AMC-traditional ARQ and that in [3] for $P_{loss}$=0.001.

In Fig. 3 we plot the spectral efficiency of the adaptive rate and fixed rate cooperative ARQ. As evident the proposed joint AMC-cooperative ARQ scheme provides much higher spectral efficiency gain when compared to fixed rate cooperative ARQ thanks to the use of CSI at the source and relay.

As specified in Corollary 1, for $d = 0$, the proposed scheme reduces to a special case of AMC combined with the traditional ARQ scheme. In [3], a similar approach for jointly designing of AMC with traditional ARQ is proposed. As depicted in Fig. 4, the spectral efficiency of the proposed AMC design with traditional ARQ outperforms that of [3], especially for smaller SNRs. This is because the proposed scheme uses different target PERs for transmission and retransmissions of a packet in an optimized manner.

## VI. Conclusions

In this paper we developed a cross-layer approach to jointly design AMC at the physical and cooperative ARQ at the data link layer to enhance the system performance for transmissions of data packet over block fading relay channels. The proposed scheme maximizes the system spectral efficiency subject to a prescribed PLR constraint for delay constrained packet services. Numerical results indicate a considerable spectral efficiency gain can be achieved in compare with AMC-only at the physical layer. Moreover, the proposed adaptive rate cooperative ARQ scheme outperforms the fixed rate cooperative ARQ when the transmission modes at the source and relay nodes are selected based on the channel statistics. This in turn validates the efficiency of the proposed cross-layer approach for designing AMC schemes in wireless relay networks.

As a future work we are investigating the benefits of joint AMC-cooperative ARQ design over correlated fading channels such as land mobile satellite channels where due to burst errors the traditional ARQ schemes degrade, severely.

ACKNOWLEDGEMENT

This work has been supported in part by the Iran Telecommunications Research Center.

## APPENDIX A

Given the instantaneous SNRs $\gamma_1$, $\gamma_2$, and a fixed SNR $\gamma_{sr}$ for the S-D, R-D, and S-R channels respectively, using the total probability theorem, the average PLR of the proposed scheme is given by

$$\overline{PLR}(\gamma_{sr}) = \iint \Pr(\text{Loss of packet}|\gamma_1,\gamma_2) p_{\gamma_1,\gamma_2}(\gamma_1,\gamma_2) d\gamma_1 d\gamma_2 \quad (16)$$

where

$$\Pr(\text{Loss of packet}|\gamma_1,\gamma_2) = \Pr(T_{sd}:f, T_{sr}:f|\gamma_1,\gamma_2) + \Pr(T_{sd}:f, T_{sr}:s, T_{rd}:f|\gamma_1,\gamma_2) \quad (17)$$

Since the channel SNRs $\gamma_1$ and $\gamma_2$ are independent, we have

$$\Pr(T_{sd}:f, T_{sr}:f|\gamma_1,\gamma_2) = PER_n(\gamma_1)\varepsilon_n \quad (18)$$
$$\Pr(T_{sd}:f, T_{sr}:s, T_{rd}:f|\gamma_1,\gamma_2) = (1-\varepsilon_n)PER_n(\gamma_1)PER_m(\gamma_2)$$

where $T_{sd}$, $T_{sr}$, $T_{rd}$ denote the transmission over S-D, S-R and R-D links, respectively. $f$ and $s$ also denote the success and failure of the transmission, respectively. In the scenario under consideration, in the outage modes of S-D (*i.e.* $\gamma_1 < \gamma_{1,1}$) and R-D links (*i.e.* $\gamma_2 < \gamma_{2,1}$), no data is transmitted by the S and R nodes. Therefore, using (16) and (17) the average system PLR can be calculated as

$$\overline{PLR}(\gamma_{sr}) = \frac{1}{\Pr(\gamma_1 > \gamma_{1,1})} \int_{\gamma_{1,1}}^{\infty} \Pr(T_{sd}:f, T_{sr}:f|\gamma_1) p_{\gamma_1}(\gamma_1) d\gamma_1 + \frac{1}{\Pr(\gamma_1 > \gamma_{1,1}, \gamma_2 > \gamma_{2,1})} \int_{\gamma_{1,1}}^{\infty}\int_{\gamma_{2,1}}^{\infty} \Pr(T_{sd}:f, T_{sr}:s, T_{rd}:f|\gamma_1,\gamma_2) \times p_{\gamma_1}(\gamma_1) p_{\gamma_2}(\gamma_2) d\gamma_1 d\gamma_2 \quad (19)$$

Substituting the equation (18) in (19) we can obtain

$$\overline{PLR}(\gamma_{sr}) = \frac{\sum_{n=1}^{N}\varepsilon_n \int_{\gamma_{1,n}}^{\gamma_{1,n+1}} PER_n(\gamma_1)p_{\gamma_1}(\gamma_1)d\gamma_1}{\sum_{n=1}^{N}\int_{\gamma_{1,n}}^{\gamma_{1,n+1}} p_{\gamma_1}(\gamma_1)d\gamma_1} \quad (20)$$
$$+ \left(\frac{\sum_{n=1}^{N}(1-\varepsilon_n)\int_{\gamma_{1,n}}^{\gamma_{1,n+1}} PER_n(\gamma_1)p_{\gamma_1}(\gamma_1)d\gamma_1}{\sum_{n=1}^{N}\int_{\gamma_{1,n}}^{\gamma_{1,n+1}} p_{\gamma_1}(\gamma_1)d\gamma_1}\right)$$
$$\times \left(\frac{\sum_{m=1}^{N}\int_{\gamma_{2,m}}^{\gamma_{2,m+1}} PER_n(\gamma_2)p_{\gamma_2}(\gamma_2)d\gamma_2}{\sum_{m=1}^{N}\int_{\gamma_{2,m}}^{\gamma_{2,m+1}} p_{\gamma_2}(\gamma_2)d\gamma_2}\right)$$

after following a few steps, the PLR in equation (20) is expressed as

$$\overline{PLR}(\gamma_{sr}) = \left(\frac{\sum_{n=1}^{N}\overline{PER}_n^{sd} P_n^{sd}}{\sum_{n=1}^{N} P_n^{sd}}\right)\left(\frac{\sum_{m=1}^{N}\overline{PER}_m^{rd} P_m^{rd}}{\sum_{m=1}^{N} P_m^{rd}}\right)$$
$$+ \left(\frac{\sum_{n=1}^{N}\varepsilon_n \overline{PER}_n^{sd} P_n^{sd}}{\sum_{n=1}^{N} P_n^{sd}}\right) \times \left(1 - \frac{\sum_{m=1}^{N}\overline{PER}_m^{rd} P_m^{rd}}{\sum_{m=1}^{N} P_m^{rd}}\right) \quad (21)$$